\documentclass[prd,preprint,eqsecnum,nofootinbib,amsmath,amssymb,preprintnumbers,
tightenlines,longbibliography]{revtex4}

\usepackage{afterpage}
\usepackage{mathrsfs}
\usepackage{graphicx}
\usepackage{bm}
\usepackage{paralist}
\usepackage[colorlinks,linkcolor=blue,citecolor=blue]{hyperref}

\def\st{\begin{equation}}
\def\stp{\end{equation}}
\def\bg{\begin{eqnarray}}
\def\nd{\end{eqnarray}}

\advance\parskip 1.9pt
\advance\voffset -0.2in

\begin{document}

\title{Longitudinal fluctuations of the fireball density in \\heavy-ion collisions} 

\author{Adam~Bzdak}
\email{abzdak@bnl.gov}
\affiliation
    {%
    RIKEN BNL Research Center,
    Brookhaven National Laboratory, \\
    Upton, NY 11973, USA
    }%

\author{Derek~Teaney}
\email{derek.teaney@stonybrook.edu}
\affiliation
    {%
    Department of Physics \& Astronomy,
    Stony Brook University,
    Stony Brook, NY 11794, USA \\\\
    }%

\begin{abstract}
We show that fluctuations of the fireball shape in the longitudinal direction
generate nontrivial rapidity correlations that depend not only on the rapidity
difference, $y_{1} - y_{2}$, but also on the rapidity sum, $y_{1} + y_{2}$. 
This is explicitly demonstrated in a simple wounded nucleon model, and  the
general case is also discussed. We show how to extract different components
of the fluctuating fireball shape from the measured two-particle rapidity
correlation function. The experimental possibility of studying the longitudinal
initial conditions in heavy-ion and proton-proton collisions is emphasized.
\end{abstract}
\maketitle

\section{Introduction}

When a fireball produced in heavy-ion collisions is studied in an azimuthal
angle, it is very useful to parametrize its initial shape with the help of the
Fourier decomposition \cite%
{Ollitrault:1992bk,Ackermann:2000tr,Voloshin:2008dg}. 
This decomposition clarifies trends in  the elliptic flow,   
the fluctuations in  elliptic flow, and the higher harmonics flows, which  are associated with the triangularity and other shape parameters of the fireball \cite%
{Alver:2010gr,Teaney:2010vd}.

The harmonic analysis gives important information about the
transverse initial conditions in heavy-ion collisions, and the
mechanism of the subsequent evolution of the produced fireball -- see {\it e.g.} \cite%
{Florkowski:book}.  In particular, the success of the hydrodynamic 
model in describing the $v_n$ data at RHIC and LHC places new  constraints
on the initial conditions in the transverse direction.

A similar idea can be applied to study the shape of the fireball in the
longitudinal direction. In this paper we will focus on rapidity, $y$, but
our arguments hold for any longitudinal variable. This study was initiated
in Ref. \cite{Bialas:2011bz} (see also \cite{Bialas:2012zz}) where it was
argued that  long-range rapidity correlations can be interpreted in terms
of the fluctuating rapidity density of the created fireball. When applied to
the STAR data \cite{Abelev:2009ag}, a significant asymmetric component in
the fireball's rapidity shape was found in the most central Au+Au
collisions.

In this paper we extend the discussion presented in Ref. \cite{Bialas:2011bz}. 
We  demonstrate that the fluctuations in the fireball rapidity
density result in a nontrivial structure of the rapidity correlation
function, and  propose to study the additional components beyond
asymmetry described above. The experimental method to
extract various components is also discussed.

The structure of this paper is following. In the next section we discuss the
problem in a simple model. We show that an event-by-event difference between
the number of wounded nucleons in the target and the projectile results in a
long-range asymmetry of the fireball. We derive the correlation function and
show that it depends  on both the rapidity difference and the rapidity sum.
In section 3, we decompose the different components of the fireball
rapidity density with Chebyshev polynomials, and show how to extract the
strength of these components from the measured rapidity correlation
function. The practical application of this idea is discussed in section 4,
where we also include several comments. We summarize our paper with the
conclusions in section 5.

\section{Simple model}

In this section we explicitly demonstrate in a simple model that an
event-by-event global fluctuations of the fireball rapidity shape lead to a
nontrivial two-particle rapidity correlation function.

For a given heavy ion event, we denote 
the number of wounded nucleons moving  to the left and to the right with $w_L$ and $w_R$  respectively.
The average over many events will be denoted by $\left\langle
w_{L}\right\rangle $ and $\left\langle w_{R}\right\rangle $. 
In collisions characterized by $\left\langle w_{L}\right\rangle \neq
\left\langle w_{R}\right\rangle $, the single particle rapidity distribution
is asymmetric with respect to $y=0$, where $y$ represents the rapidity in
the center-of-mass frame. This asymmetry is clearly evident in d+Au collisions as measured at RHIC \cite{Back:2004mr},  and is easily
reproduced by practically all models of heavy-ion collisions -- see {\it e.g.} \cite%
{Bialas:2007eg,Deng:2010mv,Tribedy:2011aa,Bozek:2011if}.

In symmetric heavy ion collisions (Au+Au for example) we have $%
\left\langle w_{L}\right\rangle =\left\langle w_{R}\right\rangle $, and the
single particle rapidity distribution is obviously symmetric with respect to 
$y=0$ \cite{Back:2001bq}. However, this distribution is symmetric only when
averaged over many events. In a single event the shape (in rapidity) of the
fireball may be asymmetric since $w_{L}\neq w_{R}$.\footnote{%
The asymmetry due to the finite number of produced particles is not 
relevant to this analysis.
} Indeed, in a single event the number of wounded nucleons going
to the left may differ from the number of wounded nucleons going to the
right. As discussed below, the asymmetry can be quantified by $\left\langle
(w_{L}-w_{R})^{2}\right\rangle $, 
which is significantly larger than zero in Au+Au collisions.

It is a useful exercise to calculate in a simple model the two-particle
rapidity correlation function originating from fluctuations in $w_{L}-w_{R}$. 
Here we consider the wounded nucleon model \cite{Bialas:1976ed}, which
is a very useful model for understanding many features of
heavy-ion data \cite%
{Bialas:2004su,Bialas:2007eg,Bzdak:2009dr}. To simplify our
considerations, let us assume that the single particle rapidity distribution
measured in d+Au collisions can be approximated by a liner function of
rapidity\footnote{%
This assumption is quite reasonable outside the fragmentation regions \cite%
{Back:2004mr}.}. Consequently the distribution from a single wounded nucleon
is also a linear function of rapidity. In the wounded nucleon model, the
single particle distribution at a given $w_{L}$ and $w_{R}$ is given by \cite%
{Bialas:2004su}%
\begin{eqnarray}
\rho (y;w_{L},w_{R}) &=&w_{R}(a+by)+w_{L}(a-by)  \nonumber \\
&=&a\left( w_{L}+w_{R}\right) -by\left( w_{L}-w_{R}\right) ,  \label{roLR}
\end{eqnarray}%
where $a+by$ is the rapidity distribution from a right-mowing wounded
nucleon, and $a-by$ is the contribution form a left-mover. As seen from
above equation we have an asymmetric component that is proportional to $y$. 
Assuming further that at a given $w_{L}$ and $w_{R}$ there are no
correlations in the system\footnote{%
We want to study correlations originating only from shape fluctuations and 
we neglect all other correlations. We will come back to this point in
section 4.}, the two-particle rapidity distribution at a given $w_{L}$ and $%
w_{R}$ is%
\begin{eqnarray}
\rho _{2}(y_{1},y_{2};w_{L},w_{R}) &=&\rho (y_{1};w_{L},w_{R})\rho
(y_{2};w_{L},w_{R})  \nonumber \\
&=&a^{2}\left( w_{L}+w_{R}\right)
^{2}-ab(w_{L}^{2}-w_{R}^{2})(y_{1}+y_{2})+y_{1}y_{2}b^{2}(w_{L}-w_{R})^{2}.
\label{ro2LR}
\end{eqnarray}%
Summing Eq. (\ref{ro2LR}) over $w_{L}$ and $w_{R}$ with an appropriate
probability distribution, $P(w_{L},w_{R})$, we obtain the experimentally
accessible two-particle rapidity distribution. Taking $\left\langle
w_{L}^{2}\right\rangle =\left\langle w_{R}^{2}\right\rangle $, corresponding
to symmetric Au+Au collisions, we obtain%
\begin{equation}
\rho _{2}(y_{1},y_{2})=a^{2}\left\langle \left( w_{L}+w_{R}\right)
^{2}\right\rangle +y_{1}y_{2}b^{2}\left\langle
(w_{L}-w_{R})^{2}\right\rangle .
\end{equation}%
Consequently, the two-particle rapidity correlation function reads 
\begin{eqnarray}
C(y_{1},y_{2}) &\equiv&\rho _{2}(y_{1},y_{2})-\rho (y_{1})\rho (y_{2})  \, ,  \nonumber
\\
&=&a^{2}\left[ \left\langle w_{+}^{2}\right\rangle -\left\langle
w_{+}\right\rangle ^{2}\right] +y_{1}y_{2}b^{2}\left\langle
w_{-}^{2}\right\rangle ,  \label{C}
\end{eqnarray}%
where $w_{+}=$ $w_{L}+w_{R}$ and $w_{-}=w_{L}-w_{R}$. As seen from Eq. (\ref%
{C}) the fluctuations in $w_{L}-w_{R}$ result in a nontrivial rapidity
structure of the two-particle correlation function. $C(y_1,y_2)$ depends not only on
the rapidity difference, $y_{-}=y_{1}-y_{2}$, but also on the rapidity sum, $%
y_{+}=y_{1}+y_{2}$. Indeed,
the correlation function 
\begin{equation}
C(y_{1},y_{2})\sim y_{1}y_{2}b^{2}\left\langle w_{-}^{2}\right\rangle =\frac{%
1}{4}b^{2}(y_{+}^{2}-y_{-}^{2})\left\langle w_{-}^{2}\right\rangle ,
\end{equation}%
decreases as a function of rapidity difference, $y_{-}$, and increases 
as a function of rapidity sum, $y_{+}$. This
dependence on $y_+$ can distinguish  fluctuations of the fireball
shape from  well known sources of correlations (such as resonance decays) that depend mainly on $y_{1}-y_{2}$.

Eq. (\ref{C}) should be taken as an illustration of the
problem we would like to adress in this paper. Despite its
simplicity, the model result shows quite convincingly that the event-by-event asymmetry of the fireball shape 
that is present in symmetric heavy-ion collisions can lead to interesting rapidity correlations \cite{Bialas:2011bz}. Obviously there can
be more complicated sources of this asymmetry in more realistic models, {\it e.g.} the difference in the
number of flux-tubes in the CGC/Glasma approach \cite{Gelis:2010nm,Dusling:2009ni}.

\begin{figure}[tbp]
\includegraphics[height=2.5in]{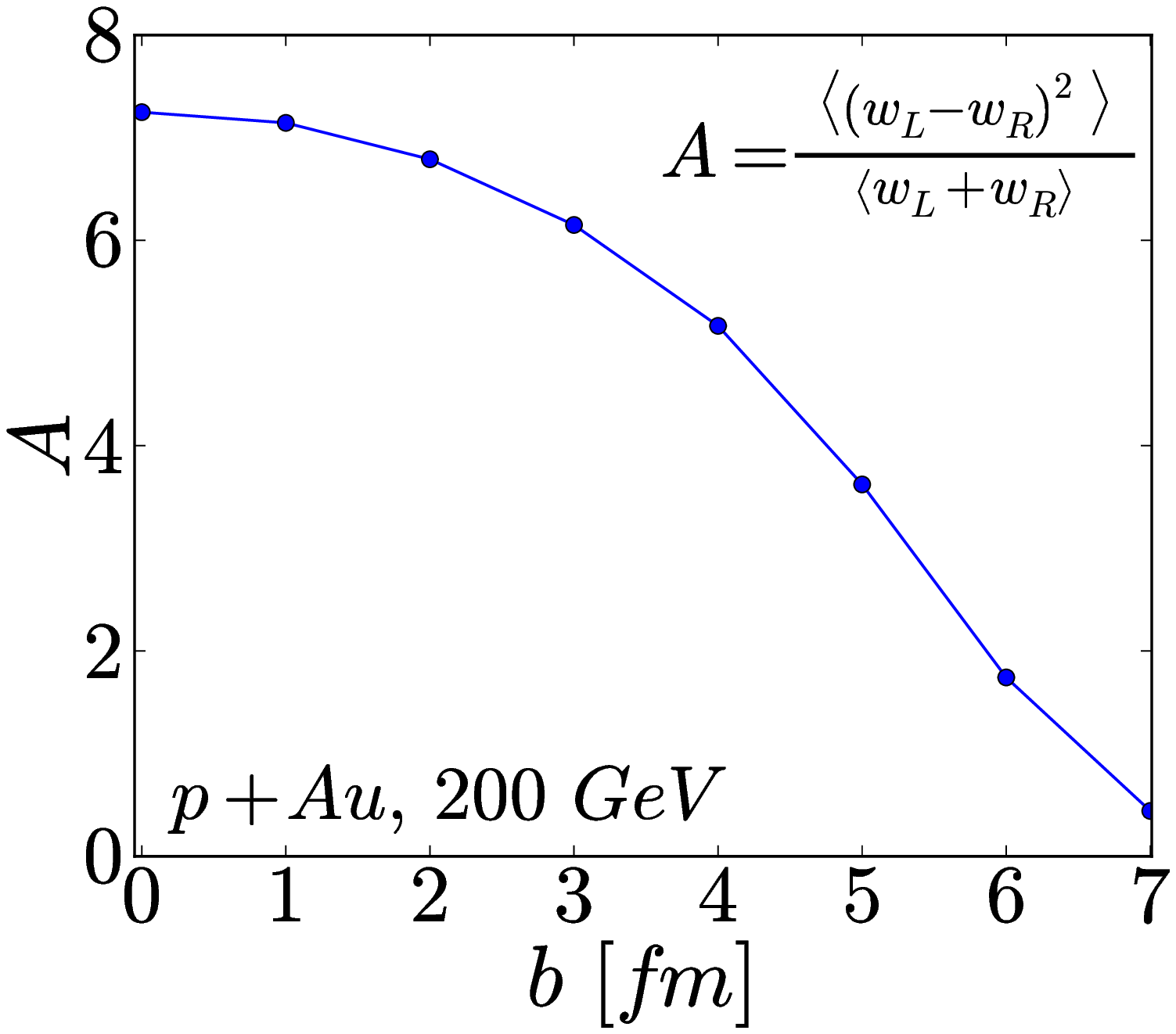} \hfill %
\includegraphics[height=2.45in]{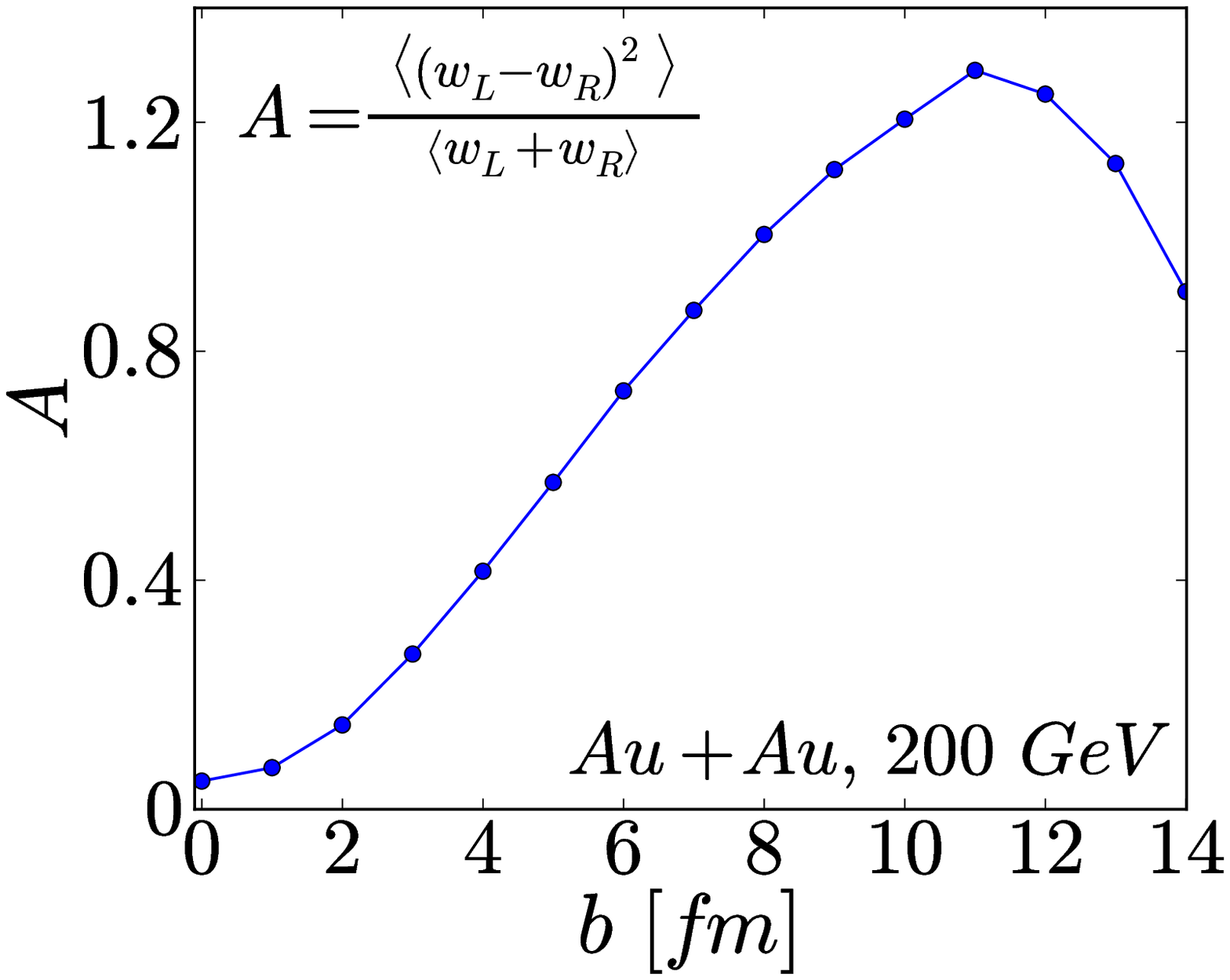}
\caption{Asymmetry in wounded nucleons in p+Au (left) and Au+Au (right)
collisions at $\protect\sqrt{s}=200$ GeV as a function of the impact
parameter $b$.}
\label{fig_A}
\end{figure}

In Fig. \ref{fig_A} we present $\left\langle (w_{L}-w_{R})^{2}\right\rangle $
divided by the total number of wounded nucleons $\left\langle
w_{L}+w_{R}\right\rangle $. We performed our calculations at $\sqrt{s}=200$
GeV in Au+Au and p+Au collisions in the Monte-Carlo Glauber model. 
It is interesting to notice that $\left\langle
(w_{L}-w_{R})^{2}\right\rangle $ in Au+Au collisions is quite large and
comparable in magnitude to the total number of wounded nucleons. Thus,
event-by-event rapidity fluctuations 
are of order $\sim 1/(w_L+w_R)^{1/2}$ compared to the average, and this
is large compared to normal statistical fluctuations of order $\sim 1/N^{1/2}$,
where $N$ is the number of particles.

In the next section we generalize  Eq. (\ref{C}) to analyze
arbitrary rapidity fluctuations of the fireball shape.

\section{General shape fluctuations}

In the previous section we discussed the asymmetric component of the
fireball rapidity density, originating from a non-zero values of $%
w_{L}-w_{R} $. As seen in Eq. (\ref{roLR})  for the simple model of
the previous section,  the single particle rapidity distribution
at a given $w_{L}-w_{R}$ is proportional to rapidity $y$,  
{\it i.e.}  the fireball is denser on one side of the rapidity window
than on the other. There may be different
sources of this asymmetry such as the left-right difference in the
number of collisions, or the difference in the number of asymmetric
long-range flux-tubes in the CGC/Glasma approach \cite{Gelis:2010nm,Dusling:2009ni}. Let us
denote the parameter that characterizes this asymmetry by $a_{1}$. In our
simple model, $a_{1}\propto w_{L}-w_{R}$.

Equation (\ref{roLR}) also contains a term that is proportional to the total
number of wounded nucleons. Fluctuations of this quantity lead to symmetric,
rapidity independent, fluctuations of the \emph{whole} fireball. This can
naturally originate from  impact parameter fluctuations, which are
always present in heavy-ion collisions. Let  $a_{0}$ denote the
parameter that characterizes this source of fluctuation. In our simple
model, $a_{0}\propto w_{L}+w_{R}$.

\begin{figure}[b]
\includegraphics[height=2.5in]{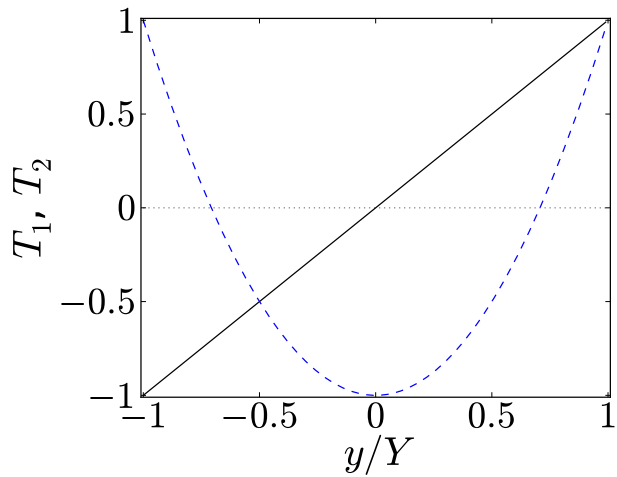}
\caption{Two components of the fluctuating fireball rapidity density: An
asymmetry (solid black line) and a butterfly (dashed blue line).}
\label{fig_cheb}
\end{figure}

The natural question arises if there are more components in the fluctuating
shape of the fireball.  For example,
a ``butterfly'' component would characterize a symmetric fireball with higer (or lower) 
density on both sides of the midrapidity region, and lower (or higher)  density at mid-rapidity.%
\footnote{%
Such  ``butterfly" fluctuations are suggested by the measured 
forward-backward rapidity correlations at RHIC \cite{Abelev:2009ag}. For a fixed number of particles at midrapidity, 
  it was observed that the particle yields in pair of narrow rapidity bins located
symmetrically about midrapidity strongly fluctuate. Surprisingly, these 
fluctuations are also strongly correlated. Thus, even if the density  is
approximately fixed in the middle of the fireball, both sides of the
fireball fluctuate together. The physical origin of this correlation is currently under investigation -- see Refs. \cite{Bzdak:2011nb,Lappi:2009vb}.}  The single particle
rapidity distribution affected by this component would be proportional to $%
y^{2}$. Let us denote by $a_{2}$ the parameter that characterizes the
strength of this effect.  We will parametrize the fireball asymmetry
and the butterfly component with the two Chebyshev polynomials, $T_1(y/Y)$ and $T_2(y/Y)$, which are shown in Fig. \ref{fig_cheb}.

Similarly, we can introduce additional components to fully parametrize shape fluctuations in rapidity. 
Thus, it is tempting to expand the single
particle rapidity distribution at a given $a_{0},a_{1},...$ in terms of the
orthogonal polynomials%
\begin{equation}
\rho (y;a_{0},a_{1},...)=\rho (y)\left[ 1+\sum\nolimits_{i=0}a_{i}T_{i}%
\left( y/Y\right) \right] ,  \label{ro-an}
\end{equation}%
where $\rho (y)$ is the single particle distribution averaged over $%
a_{0},a_{1},\ldots$\,. Here we have expanded the distribution in Chebyshev polynomials\footnote{%
For example: $T_{0}(x)=1,$ $T_{1}(x)=x,$ $T_{2}(x)=2x^{2}-1,$ $%
T_{3}(x)=4x^{3}-3x$ etc.}, but other choices are certainly possible. 
The parameter $Y$ characterizes the
scale of long-range rapidity fluctuations  in the system. We will discuss reasonable values of $Y$ in the next section. Averaging both
sides of Eq. (\ref{ro-an}) over $a_{0},a_{1},...$ with an appropriate
probability distribution, $P(a_{0},a_{1},...)$, we obtain $\left\langle
a_{i}\right\rangle =0$ for all $i\geq 0$.

Assuming that at a given $a_{0},a_{1},...$ there are no other large sources of long-range rapidity correlations, the two-particle rapidity distribution is%
\begin{equation}
\rho _{2}(y_{1},y_{2};a_{0},a_{1},...)=\rho (y_{1};a_{0},a_{1},...)\rho
(y_{2};a_{0},a_{1},...) \, .
\end{equation}%
Taking an average over $a_{i}$ and subtracting $\rho (y_{1})\rho (y_{2})$,
we obtain the  two-particle rapidity correlation function%
\begin{equation}
C(y_{1},y_{2})=\rho (y_{1})\rho (y_{2})\left[ \sum\nolimits_{i,k=0}\left%
\langle a_{i}a_{k}\right\rangle T_{i}\left( y_{1}/Y\right) T_{k}\left(
y_{2}/Y\right) \right] \, .  \label{C-gen}
\end{equation}%
It is useful to recall the physical meaning of the first few terms in Eq. (\ref%
{C-gen}): $\left\langle a_{0}^{2}\right\rangle $ represents the rapidity
independent fluctuations of the fireball as a whole, 
$\left\langle
a_{0}a_{1}\right\rangle y_{2}$ describes the correlation between rapidity
independent fluctuations of the fireball and its asymmetry, $\left\langle
a_{1}^{2}\right\rangle y_{1}y_{2}$ is the asymmetric component discussed in
the previous section,  and $\left\langle a_{2}^{2}\right\rangle [2\left(
y_{1}/Y\right) ^{2}-1][2\left( y_{2}/Y\right) ^{2}-1]$ represents the
butterfly contribution, {\it etc.}\,.

From the previous section, we know that the asymmetric component, $%
\left\langle a_{1}^{2}\right\rangle $, introduces a long-range rapidity
correlation that is a decreasing  function of the rapidity difference, $%
y_{-}=y_{1}-y_{2}$, and an increasing function of the rapidity sum, $%
y_{+}=y_{1}+y_{2}$. It is a straightforward  to verify that the
rapidity structure originating from the butterfly component leads to a
correlation function that is decreasing both in $y_{-}$ and $y_{+}$.

To conclude this section,  we  point out that the values of $%
\left\langle a_{i}a_{k}\right\rangle $ can be extracted directly from the
correlation function $C(y_{1},y_{2})$. Using the orthogonality of Chebyshev polynomials%
\begin{equation}
\int_{-1}^{1}T_{i}(x)T_{k}(x)\left( 1-x^{2}\right) ^{-1/2}dx=c_{i}\delta
_{i,k} \, ,
\end{equation}%
where $c_{0}=\pi $ and $c_{i}=$ $\pi /2$ for $i>0$, we obtain 
\begin{equation}
\left\langle a_{i}a_{k}\right\rangle =\frac{1}{c_{i}c_{k}}\int_{-Y}^{Y}\frac{%
C(y_{1},y_{2})}{\rho (y_{1})\rho (y_{2})}\frac{T_{i}(y_{1}/Y)T_{k}(y_{2}/Y)}{%
\left[ 1-(y_{1}/Y)^{2}\right] ^{1/2}\left[ 1-(y_{2}/Y)^{2}\right] ^{1/2}}%
\frac{dy_{1}dy_{2}}{Y^{2}} \, .  \label{aiak}
\end{equation}
In the next section we will discuss how Eq. (%
\ref{aiak}) can be used in practice.

\section{Comments}

In this section we list several comments to clarify the
analysis presented in this paper.
In deriving  Eq. (\ref{C-gen}) and Eq. (\ref{C}), we assumed
that at a given $a_{0},a_{1},...$ there are no correlations in the system.
In other words, the only sources of correlations are fluctuations in the
fireball rapidity density. Unfortunately, short-range correlations may
contribute to the left-hand side of Eq. (\ref{C-gen}), and contaminate the
signal coming from the shape fluctuations. Particularly problematic may be
the correlations from resonance decays and local local charge conservation \cite%
{Schlichting:2010qia,Bozek:2012en}. These problems can be mitigated  by studying Eq. (\ref%
{C-gen}) for positive and negative particles separately,  which significantly
reduces these unwanted backrounds. Moreover, the dependence of the correlation
function on the rapidity sum, $y_{+}$, can be used to distinguish
between rapidity density fluctuations and the 
short-range correlations of the background.

One could also worry that at a given $w_{L}+w_{R}$ the distribution
of final particles is given approximately  by a negative binomial
distribution (NBD) \cite{Adare:2008ns,Gelis:2009wh} which  introduces long-range rapidity correlations into the system. This concern
is unfounded, however, because
the NBD leads to the following two-particle rapidity distribution\footnote{%
We sample particles from NBD and distribute them randomly in rapidity
according to $\rho (y)$.}%
\begin{equation}
\rho _{2}(y_{1},y_{2})=\rho (y_{1})\rho (y_{2})\left( 1+1/k\right) ,
\end{equation}%
where $k$ measures deviation from Poisson distribution. As seen from Eq. (%
\ref{C-gen}), the NBD $\rho _{2}$ influences only $\left\langle
a_{0}^{2}\right\rangle $. In fact, this is the expected result, since the NBD can be
viewed as a rapidity independent fluctuation of the whole fireball.

As pointed out in the previous section, it is not obvious what is
the appropriate value of $Y$ in the preceding formulas. Clearly,
$Y$ parametrizes the range of global rapidity fluctuations in
the fireball density.
For instance, at $\sqrt{s}=200$ GeV the single particle distribution in
d+Au collisions \cite{Back:2004mr} is approximately linear as a function of $%
y$ for $\left| y\right| <2$. Thus, for the asymmetric component parametrized by $a_{1}$, $Y\approx 2$ is a reasonable choice. For higher and
lower  energies, this parameter can be  rescaled by the ratio of 
beam rapidities. This value of $Y$ roughly corresponds  to the size of
the thermal fireball, and it is plausible that higher components, if
they exist, are present in this region. If the measurement is performed in
the smaller window than $[-Y,Y]$,  fitting the measured correlation
function with Eq. (\ref{C-gen}) can determine the parameters of interest, $%
\left\langle a_{i}a_{k}\right\rangle $.

Another choice is to assign $Y$ to 
the rapidity interval of the measurement, and to 
investigate how the components $\left\langle a_i a_k\right\rangle$ change when
this rapidity scale is varied.
It is possible that the different fluctuating components are visible at different rapidity scales, and a systematic study of this sort can sort out these differences.

It would be very interesting to compare the strengths of different components,
$\left\langle a_{i}a_{k}\right\rangle $, between heavy-ion and proton-proton
collisions. This could reveal interesting differences in the longitudinal
initial conditions between these two systems. For instance, in Ref.
\cite{Bialas:2011bz} it was shown that the asymmetric component is
significantly stronger in p+p collisions than in central Au+Au collisions.

Further, the ideas presented in this paper could be extended by incorporating
the multi-bin analysis proposed in Refs. \cite{Bzdak:2009bc,Bialas:2010zb,Bialas:2011xk}.
This multi-bin analysis can be used 
to investigate the different sources of particles production, providing
a detailed picuture of the fireball in the longitudinal direction.
Finally, we point out that the results obtained in this paper
can be easily generalized to three- and many-particle correlation functions.

\section{Conclusions}

In conclusion, we showed that event-by-event fluctuations of the fireball
rapidity density introduce interesting rapidity correlations that depend
both on the rapidity difference, $y_{1}-y_{2},$ and the rapidity sum, $%
y_{1}+y_{2}$. We demonstrated this explicitly in the wounded nucleon model,
where an event-by-event difference between the number of wounded nucleons in
a target and a projectile, $w_{L}-w_{R}$, leads to the long-range asymmetry
of the fireball. The resulting correlation function in symmetric A+A
collisions is given in Eq. (\ref{C}).

We further proposed to expand the measured two-particle rapidity correlation
function in a series of the Chebyshev polynomials (see Eq. (\ref{C-gen})), 
where each polynomial represents a different component of the fireball's
fluctuating rapidity density. The quadratic polynomial in this expansion
describes the ``butterfly'' fluctuations described above,  which are suggested by recent measurements at
RHIC. The coefficients of this expansion, $\left\langle
a_{i}a_{k}\right\rangle $, characterize the strength of various components,
and we propose to extract these coefficients  from the measured correlation
function. This can reveal nontrivial information about the structure of the
fireball in the longitudinal direction, and can test various models of particle
production in hadronic collisions.

\vspace{\baselineskip}
\noindent{ \bf Acknowledgments:} \\
{}\\
We thank Andrzej Bialas and Larry McLerran for interesting discussions and encouragement. 
A.~Bzdak is supported through the RIKEN-BNL Research Center, and by a grant from the Polish
Ministry of Science and Higher Education, No. N202 125437.
D.~Teaney is a RIKEN-RBRC research fellow, and is supported  by the
Sloan Foundation and by the Department of Energy through the Outstanding Junior
Investigator program, DE-FG-02-08ER4154.

\end{document}